\documentclass[twocolumn,showpacs,preprintnumbers,amsmath,amssymb]{revtex4}
\usepackage{graphicx}
\usepackage{dcolumn}
\usepackage{bm}
\usepackage{epsfig}
\usepackage{amsfonts}

\begin{document}

\title{Superstatistical distributions from a maximum entropy principle}

\author{Erik Van der Straeten}

\author{Christian Beck}
\affiliation{School of Mathematical Sciences, Queen Mary,
University of London, Mile End Road, London E1 4NS, UK}
\email[E-mail:\ ]{e.straeten@qmul.ac.uk, c.beck@qmul.ac.uk}

\date{\today}

\begin{abstract}
We deal with a generalized statistical description of nonequilibrium complex systems
based on least biased distributions given some prior information.
A maximum entropy principle is introduced
that allows for the determination of the distribution
of the fluctuating intensive parameter $\beta$ of a
superstatistical system, given certain constraints on the complex
system under consideration. We
apply the theory to three examples: The superstatistical quantum
mechanical harmonic oscillator, the superstatistical classical
ideal gas, and velocity time series as measured in a turbulent
Taylor-Couette flow.
\end{abstract}

\maketitle

\section{Introduction}
Many complex systems in physics, biology, medicine, and economics
exhibit a spatio-temporally inhomogeneous dynamics that can be
effectively described by a superposition of several statistics on
different time scales, in short a 'superstatistics'
\cite{beck-cohen,unicla,touchette,supergen,souza,chavanis,plastino,rajagopal,jizba,
wilk,prl01,sattin}. 
The concept of such a superposition of statistics was first systematically discussed
in
\cite{beck-cohen}, in the mean time many applications for a
variety of complex systems have been pointed out
\cite{daniels,maya,reynolds,RMT,porpo,rapisarda,cosmic,eco}.
Essential for this approach is the existence of an intensive
parameter $\beta$ that fluctuates on a much larger time scale
than the typical relaxation time of the local dynamics. In a thermodynamic
setting, $\beta$ can be interpreted as a local inverse
temperature of the system, but much broader interpretations are
possible. Locally, the system is described by equilibrium
statistical mechanics with inverse temperature $\beta$, whereas
globally there is yet another statistics of the inverse
temperature $\beta$. The two effects produce a superposition of
two statistics, or in a short, a `superstatistics'. Related statistical tools
play of course an important role in the theory of stochastic
processes, see
e.g. \cite{vKa,heston,kubo,feller}.

The stationary distributions of superstatistical systems,
obtained by averaging over all $\beta$, typically exhibit
non-Gaussian behavior with fat tails, which can decay e.g. with
a power law, or as a stretched exponential, or in a more
complicated way \cite{touchette}. In general, the
superstatistical parameter $\beta$ need not to be an inverse
temperature but can also be interpreted as an effective friction
constant, a changing mass parameter, a changing amplitude of
Gaussian white noise, the fluctuating energy dissipation in
turbulent flows, a fluctuating volatility in finance, an
environmental parameter for biological systems, or simply a local
variance parameter extracted from a given experimental time
series. Recent applications of the concept include hydrodynamic turbulence \cite{prl,beck03,reynolds,unicla}, pattern forming systems \cite{daniels}, cosmic rays \cite{cosmic}, solar flares \cite{maya}, share price fluctuations \cite{bouchard,ausloos,eco,eco2}, random matrix theory \cite{RMT,abul-magd2},
random networks \cite{abe-thurner}, multiplicative-noise stochastic processes \cite{queiros}, quantum
systems at low temperatures \cite{rajagopal}, wind velocity
fluctuations \cite{rapisarda}, hydro-climatic fluctuations \cite{porpo}, the statistics of train departure delays \cite{briggs} and models of
the metastatic cascade in cancerous systems \cite{chen}.

In equilibrium statistical mechanics it is clear how to obtain the
relevant probability distributions describing the long-term
behavior of the system under consideration. These are the
canonical distributions and they follow from a maximum entropy
principle. However, superstatistical systems are nonequilibrium
systems with a stationary state which is a mixture of canonical
distributions. It is a priori not clear how to obtain the mixing
distribution of the fluctuating parameter from first
principles. A promising idea to tackle this problem is to develop a more general type
of thermodynamics for superstatistical systems which leads to a
generalized maximum entropy principle that fixes these
distributions. Early attempts in this direction were made by
Tsallis and Souza \cite{souza} and later by Abe et al \cite{abc},
Crooks \cite{crooks} and Naudts \cite{naudts}. Inspired by these
early considerations, in this paper we
develop a generalized formalism that is
a) conceptually simple b) applicable to
both, classical and quantum systems c) consistent
with experimental observations. As a result, we obtain a statistical theory that can be applied to a large variety of
complex systems and which further develops the earlier ideas of Abe,
Beck, Cohen, Crooks and Naudts.

This paper is organized as follows. In section 2 we clarify
our notation and recall the
basic concept of time scale separation that lies at the heart
of any superstatistical description. In section 3 we introduce
our generalized maximum entropy principle and discuss the relation between our formalism and the previous approaches of Abe, Beck, Cohen, Crooks and
Naudts. In section 4
we discuss some physically relevant conditions on the relevant class
of probability densities. In the following sections we apply our
theory to three important examples: The superstatistical quantum
mechanical harmonic oscillator (section 5), the superstatistical
ideal gas (section 6) and velocity fluctuations as observed in
a turbulent time series (section 7). Our concluding remarks are
given in section 8.

\section{Basic concepts}
The crucial assumption of superstatistics is that the statistical
description of certain classes of complex nonequilibrium systems can be split into
two levels that have a large time scale separation. The total
system is divided into spatial cells, each in local equilibrium, but the
temperatures of the different cells don't have to be equal. As a
consequence, in very good approximation the
local properties of the different cells can be described using the standard
Boltzmann-Gibbs formalism. The main problem is then the determination
of the distribution of the temperature at the higher level of the
total nonequilibrium system. Clearly, the Boltzmann-Gibbs
formalism is not applicable at this level.

Locally, in each cell the average $\langle A\rangle_H$ of an observable $A$ is
calculated with respect to the Boltzmann-Gibbs probability measure
\begin{eqnarray}
p(H;\beta)=\frac1{Z(\beta)}e^{-\beta
H},
\end{eqnarray}
where $\beta$ is the inverse temperature, $H$ is the Hamiltonian that
describes the properties
of each spatial cell of the system, and $Z(\beta)$ is the partition function. In classical
statistical mechanics, $p(H;\beta)$ is a probability distribution
and the local average $\langle A\rangle_H$ is defined by
\begin{eqnarray}
\langle A\rangle_H&=&\int d\Gamma p(H;\beta)A,
\end{eqnarray}
with $\Gamma$ being the phase space. In quantum statistical mechanics,
$p(H;\beta)$ is a density operator and the local average $\langle
A\rangle_H$ is defined by
\begin{eqnarray}
\langle A\rangle_H&=&\textrm{Tr}p(H;\beta)A,
\end{eqnarray}
with $H$ and $A$ being operators acting on the corresponding
Hilbert space. We introduce the following shorthand notation for
the local energy $E(\beta)$ and local entropy $S(\beta)$
\begin{eqnarray}\label{ES}
E(\beta)=\langle H\rangle_H&\textrm{and}& S(\beta)=-\langle\ln
p(H;\beta)\rangle_H.
\end{eqnarray}
From a thermodynamic point of view, the Hamiltonian is an observable and the temperature is the
corresponding control parameter (intensive variable). By measurement of the average value of the observable
one can estimate the value of the corresponding control parameter. We are interested in the statistical
average of an observable $A$ of the total nonequilibrium system which has
a different inverse temperature in each cell. For this global average we will use following notation
\begin{eqnarray}
\langle\langle A\rangle_H\rangle_\beta&=&\int_0^\infty d\beta
f(\beta;\lambda_i)\langle A\rangle_H.
\end{eqnarray}
Here $f(\beta;\lambda_i)$ is the probability density
of $\beta$ in the various spatial cells, which depends on a set of parameters
$\{\lambda_i\}$ (in our notation we suppress the brackets $\{\}$).
The parameters $\lambda_i$ can be interpreted as the control
parameters corresponding with some measurable nonequilibrium
observables. Our goal in the following is to find a general principle
for the determination of $f(\beta;\lambda_i)$, given certain information that we
have on the complex system.

\section{Maximum entropy}
Let us first recall the maximum entropy principle for equilibrium
statistical mechanics, after that we will proceed to the
superstatistical extension. An impressive amount of experimental
results shows that assuming the Boltzmann-Gibbs distribution as
the equilibrium distribution of a system is a very good
approximation. Information theory gives a deeper understanding to
this success \cite{jaynes}. Usually, the only experimental
information that is available about a system is the average value
of some observables. Therefore, it is natural to use the least
biased distribution, given this prior information, as the
equilibrium distribution of the system. The practical tool to
obtain this least biased distribution is the maximum entropy
principle. Every observable that one can measure is interpreted
as a constraint. Then one introduces Lagrange multipliers and
maximizes the entropy (or negative information) under these
constraints. Using the laws of thermodynamics, one shows that the
Lagrange multipliers are related to the thermodynamic control
parameters. When one uses only the constraint that the average energy of
the system has to take on a certain value, one ends up with the
Boltzmann-Gibbs canonical distribution.

We will now extend these considerations
and use the maximum entropy principle to obtain the least
biased distribution for $f(\beta;\lambda_i)$. 
As a likelihood estimator we use the Shannon entropy, though in
principle other choices such as the Tsallis entropy \cite{tsallis}
are possible as well.
The entropy
associated with the distribution $f$ is
\begin{eqnarray}
S(\lambda_i)&=&-\langle\ln f(\beta;\lambda_i)\rangle_\beta.
\end{eqnarray}
Clearly the distribution $f(\beta;\lambda_i)$ has to be
normalized. So a first property of the distribution
$f(\beta;\lambda_i)$ that one has take into account is $\langle
1\rangle_\beta=1$. Given some complex system in a stationary
nonequilibrium state one may have additional information on the
system which imposes some additional constraints. To obtain
appropriate constraints for superstatistical systems, we briefly
repeat the general idea of this theory. In each cell, the value
of the temperature is fixed. For the entire nonequilibrium system
this condition is relaxed and the temperature is allowed to vary
between the different cells. The crucial assumption of
superstatistics is that these temperature fluctuations have a
slow time scale compared with the time scale of relaxation to
local equilibrium. The slow fluctuations of the temperature cause extra (slow)
fluctuations of variables like the entropy and the energy
in each cell. So it is reasonable to constrain that the averages of these variables
should take on certain values. One can still add further
constraints in terms of some function $g(\beta)$,
whose precise form depends on the nature of the complex system
considered, i.e. its dynamics, symmetries, and boundary conditions. Thus, in the most general case the quantity to be
optimized is
\begin{eqnarray}
&&S(\lambda_i)-\frac{\lambda_1}V\langle
S(\beta)\rangle_\beta-\frac{\lambda_2}V\langle \beta
E(\beta)\rangle_\beta
\cr
&&-\lambda_3\langle
g(\beta)\rangle_\beta-\lambda_4\langle1\rangle_\beta,
\end{eqnarray}
with $V$ being an arbitrary
constant (taking out a common factor out of the
definition of $\lambda_1$ and $\lambda_2$ will turn out to be
useful in the following). Using the well-known formula $S(\beta)=\ln Z(\beta)+\beta
E(\beta)$ and renaming
$(\lambda_1+\lambda_2)\rightarrow\lambda_2$ one obtains
\begin{eqnarray}\label{the_max}
&&S(\lambda_i)-\frac{\lambda_1}{V}\langle \ln
Z(\beta)\rangle_\beta-\frac{\lambda_2}{V}\langle \beta
E(\beta)\rangle_\beta
\cr
&&-\lambda_3\langle
g(\beta)\rangle_\beta-\lambda_4\langle1\rangle_\beta.
\end{eqnarray}
The optimization of this expression results in the following
distribution
\begin{eqnarray}\label{the_distr}
f(\beta;\lambda_i)&=&\frac{Z(\beta)^{-\lambda_1/V}}{Z(\lambda_i)}\exp\left(-\beta\lambda_2\frac{E(\beta)}{V}-\lambda_3
g(\beta)\right)
\nonumber\\
&&
\end{eqnarray}
with $Z(\lambda_i)$ a normalization constant that is fixed by the condition $\langle 1\rangle_\beta=1$.

We now relate our general result (\ref{the_distr}) to previous work obtained in the literature.
In \cite{abc}, the authors maximize
the sum of $S(\lambda_i)$ and $\langle S(\beta)\rangle_\beta$ under the constraint
of the normalization of $f(\beta;\lambda_i)$ only. This coincides
with our approach in case the Lagrange multipliers of expression
(\ref{the_max}) are chosen in the following way:
$\lambda_1/V=\lambda_2/V=-1$ and $\lambda_3=0$. This
results in a distribution that is usually not normalizable.
For this reason in \cite{abc}
the domain of $\beta$ is restricted to a
finite range when simple examples are studied, such as $n$
non-interacting classical Brownian particles. Closely related
is also the research of Crooks \cite{crooks}. He studies
general nonequilibrium systems, without assuming that the system
can be divided into different cells that reach local equilibrium.
Crooks advocates that instead of trying to obtain the probability
distribution of the entire nonequilibrium system one has to try to estimate the
'metaprobability', the probability of the microstate probability
distribution.
Crooks also uses a maximum entropy principle but puts $\lambda_3=0$.
A main difference is that Crooks does not assume local equilibrium
in the cells, hence his approach, though an interesting theoretical
construction, does not give a straightforward physical interpretation to
the fluctuating parameter $\beta$. The
advantage of our approach is that one obtains a local fluctuating
temperature that coincides with the thermodynamic temperature and
that can in principle be measured. The work of
Crooks is used by Naudts \cite{naudts} to describe equilibrium systems.
The author shows that some well-known results of equilibrium
statistical mechanics can be reformulated in a very general
context with the use of the concepts introduced in \cite{beck-cohen, crooks}.

\section{Physically relevant distributions}
We now discuss some physical properties that should
be satisfied by the distribution coming out of the entropy maximization procedure.
Physically one would expect the superstatistical
distribution $f(\beta;\lambda_i)$ to vanish at very low and very high
temperatures. Assume for the moment that no additional constraint exists,
i.e. $g(\beta)=0$. In this case
one can immediately obtain the sign of the various
Lagrange multipliers by studying the limiting behavior of $f(\beta;\lambda_i)$ for $\beta \to 0$
and $\beta \to \infty$.
In the high temperature limit, the distribution is
proportional to
\begin{eqnarray}
\frac{Z(\beta)^{-\lambda_1/V}}{Z(\lambda_i)}.
\end{eqnarray}
The partition function $Z(\beta)$  usually diverges at high temperatures (the
entropy becomes infinite). As a consequence, for physical
reasons, the sign of
$\lambda_1$ must be positive. In the low temperature limit, the
energy and the entropy go to a constant,
$\lim_{\beta\rightarrow\infty}S(\beta)=S_0$ and
$\lim_{\beta\rightarrow\infty}E(\beta)=E_0$. In this limit, the
distribution is proportional to
\begin{eqnarray}
\frac1{Z(\lambda_i)}\exp\left(-\lambda_1\frac{S_0}V-\beta\left(\lambda_2-\lambda_1\right)\frac{E_0}V\right).
\end{eqnarray}
Therefore, the sign of $\left(\lambda_2-\lambda_1\right)E_0/V$ must be
positive. Clearly, when a non-trivial function $g(\beta)\not= 0$ is implemented,
one has to take into account the limiting behavior of this function
as well. For a lot of models $E_0=0$. In these cases the temperature dependence of $\lim_{\beta\rightarrow\infty} f(\beta;\lambda_i)$ is solely determined by $g(\beta)$. This shows that implementing a non-trivial function $g(\beta)\not= 0$ as an extra constraint can be important.

Our reasoning assumes that the low temperature limits of
$S(\beta)$ and $E(\beta)$ are finite constants. This is generally true,
and is known as the third law of thermodynamics, but this limit
is only taken care of in an appropriate way if one uses quantum statistical mechanics. For
example, it is well known that the entropy of the classical ideal
gas diverges at low temperatures. Therefore we will now
illustrate the general theory with
two examples, the {\it quantum} harmonic oscillator and the {\it
classical} ideal gas. We will come back to the issue of the low
temperature limit when we study the classical ideal gas.

\section{Superstatistical quantum harmonic oscillator}
\begin{figure}
\begin{center}
\includegraphics[width=0.48\textwidth]{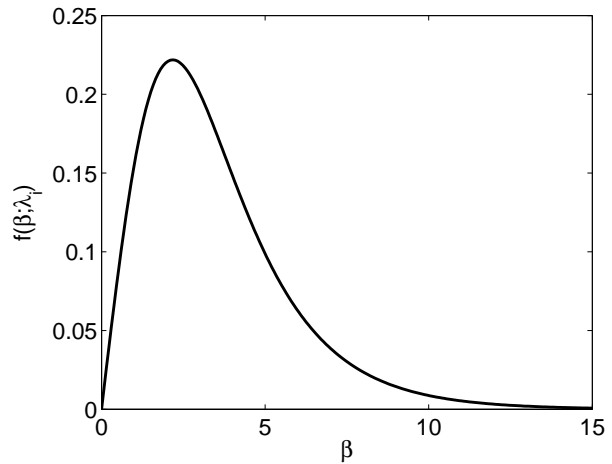}
\caption{\label{fig:plot_harm_osci}Plot of the distribution of
the inverse temperature obtained for a set of non-interacting harmonic
oscillators. The values of the parameters are $\hbar\omega=1$,
$\lambda_2=2$ and $\lambda_1=1$.}
\end{center}
\end{figure}

As a first example we study $n$ 1-dimensional non-interacting quantum
harmonic oscillators with temperature fluctuations. The Hamiltonian of a single
oscillator with mass $m$ and frequency $\omega$ is
\begin{eqnarray}
H&=&\frac1{2m}p^2+\frac12m\omega^2x^2,
\end{eqnarray}
with $p$ the momentum operator and $x$ the position operator. The
energy levels of the oscillator are well-known to be
\begin{eqnarray}
E_i&=&\hbar\omega\left(\frac12+i\right),
\end{eqnarray}
with $i=0,1,2,\ldots$. The partition function and the energy of
the $n$ oscillators become
\begin{eqnarray}
Z(\beta)&=&\left(e^{\hbar\omega\beta/2}-e^{-\hbar\omega\beta/2}\right)^{-n}
\cr
E(\beta)&=&n\hbar\omega\left(\frac12+\frac1{e^{\hbar\omega\beta}-1}\right).
\end{eqnarray}
Inserting these formulas into the expression for the distribution
of the inverse temperature (\ref{the_distr}) results in
\begin{eqnarray}\label{distr_harm_osci}
f(\beta;\lambda_i)&=&\frac{e^{\hbar\omega\beta(\lambda_1-\lambda_2)/2}}{Z(\lambda_i)\left(1-e^{-\hbar\omega\beta}\right)^{-\lambda_1}}\exp\left(-\frac{\hbar\omega\beta\lambda_2}{e^{\hbar\omega\beta}-1}\right)
\cr &&
\end{eqnarray}
with $\lambda_3=0$ and $V=n$. The high and low temperature behavior of this distribution is
\begin{eqnarray}
\lim_{\beta\rightarrow0}f(\beta;\lambda_i)&\sim&(\hbar\omega\beta)^{\lambda_1}
\cr
\lim_{\beta\rightarrow\infty}f(\beta;\lambda_i)&\sim&e^{\hbar\omega\beta(\lambda_1-\lambda_2)/2}.
\end{eqnarray}
Clearly, the distribution $f(\beta;\lambda_i)$ vanishes at high and low temperatures when $\lambda_2>\lambda_1>0$. For this quantum mechanical example, the low temperature limit of the energy $E_0=\lim_{\beta\rightarrow\infty}E(\beta)=\hbar\omega/2$ is a finite constant. As a consequence, no extra constraint ($\lambda_3=0$) is necessary to obtain a physical relevant distribution. The distribution $f(\beta;\lambda_i)$ is plotted in Fig.~\ref{fig:plot_harm_osci} for the example $\lambda_2=2$ and $\lambda_1=1$.

\section{Superstatistical classical ideal gas}
As a second example we study a 3-dimensional classical ideal gas.
The gas consists of $n$ particles with mass $m$ and is enclosed
in a box with a volume equal to unity. The partition function and
the energy of the ideal gas are
\begin{eqnarray}
Z(\beta)=\left(\frac{2\pi
m}\beta\right)^{3n/2}&\textrm{and}&E(\beta)=\frac32\frac n\beta.
\end{eqnarray}
Inserting these formulas into the expression for the distribution
(\ref{the_distr}) results in
\begin{eqnarray}\label{distr_ide_gas}
f(\beta;\lambda_i)&=&\frac{\beta^{3n\lambda_1/2V}}{Z(\lambda_i)}\exp\left(-\lambda_3
g(\beta)\right)
\end{eqnarray}
The special case $\lambda_1/V=-1$ and $\lambda_3=0$ was
already studied in \cite{abc}. As mentioned before, in that case
the distribution (\ref{distr_ide_gas}) is not normalizable and
one has to restrict the values of $\beta$ to a finite range. In
\cite{naudts} the author noticed that an inverse gamma
distribution is obtained for the choice  $\lambda_1/V=-1$,
$g(\beta)=E(\beta)$ and $\lambda_3>0$.

Let us now comment on physically reasonable choices of the
function $g(\beta)$. On physical grounds, in the various
experimental applications of the superstatistics concept so far
\cite{prl,briggs,daniels, porpo,chen,abul-magd2}, essentially
three relevant distributions $f(\beta;\lambda_i)$ were observed
for examples described by the superstatistical classical ideal
gas: The gamma distribution, the inverse gamma distribution and
the lognormal distribution. Some theoretical reasoning can be
given \cite{unicla} why this is so and why the above three
distributions span up three relevant universality classes. It is
now interesting to see that our generalized maximum entropy
principle, in contrast to previous theoretical work, contains all
these physically relevant cases. Depending on the choice of the
function $g(\beta)$ and the values of the Langrange multipliers
$\lambda_i$ one can extract the three relevant universal distributions
out off expression (\ref{distr_ide_gas}). For convenience, we put
$V=3n/2$. The gamma distribution is obtained for $g(\beta)=\beta$,
$\lambda_1>0$ and $\lambda_3>0$:
\begin{eqnarray}
f(\beta;\lambda_i)&=&\frac{\beta^{|\lambda_1|}}{Z(\lambda_i)}\exp\left(-\beta|\lambda_3|\right).
\label{di1}
\end{eqnarray}
The inverse gamma distribution is obtained for $g(\beta)=1/\beta$, $\lambda_1<0$ and $\lambda_3>0$:
\begin{eqnarray}
f(\beta;\lambda_i)&=&\frac{\beta^{-|\lambda_1|}}{Z(\lambda_i)}\exp\left(-\frac{|\lambda_3|}\beta\right).
\label{di2}
\end{eqnarray}
The lognormal distribution is obtained for $g(\beta)=(\ln\beta)^2$ and $\lambda_3>0$:
\begin{eqnarray}\label{loglam}
f(\beta;\lambda_i)&=&\frac{\beta^{\lambda_1}}{Z(\lambda_i)}\exp\left(-|\lambda_3|(\ln\beta)^2\right)
\cr
&=&\frac1{Z'(\lambda_i)}\frac1\beta\exp\left(-|\lambda_3|\left(\ln\beta-\lambda_4\right)^2\right),
\label{di3}
\end{eqnarray}
with
\begin{eqnarray}
\lambda_4&=&\frac1{2|\lambda_3|}\left(\lambda_1+1\right).
\end{eqnarray}
Unlike the quantum mechanical case, for classical complex systems
usually $g(\beta)\not= 0$ is needed to make expectations formed
with $f(\beta;\lambda_i)$ converge. This function $g(\beta)$ is
determined by additional information that one has on the complex
system under consideration (an example will be treated in the
next section).

Unlike the quantum mechanical case, for the
classical ideal gas one has to be careful
with a range of $\beta$ that goes from $0$ to $\infty$. For
this example the limiting behavior of the energy and the entropy
at low temperatures is
\begin{eqnarray}
 \lim_{\beta\rightarrow\infty}E(\beta)=0, &\textrm{and}&\lim_{\beta\rightarrow\infty}S(\beta)=-\infty.
\end{eqnarray}
The limit of the energy is acceptable from a thermodynamical
point of view, the limit of the entropy is not.
Clearly, the problem arises from the fact that the classical
treatment of an ideal gas in the low-temperature limit does not make
sense, one certainly has to take quantum corrections into
account. However, when $f(\beta;\lambda_i)$ is vanishing in
this limit, the contribution of the quantum region to the average
values of the observables will be negligible. Notice that the three aforementioned distributions 
(\ref{di1}), (\ref{di2}), (\ref{di3}) all have a single
peak at a well defined temperature. So as long as this single peak is situated in the classical region one can use classical models in the
context of superstatistics although one has to be careful in evaluating
the low temperature behavior of $f(\beta;\lambda_i)$ itself.

\section{Turbulent Taylor-Couette flow}
\begin{figure}
\begin{center}
\parbox{0.48\textwidth}{\includegraphics[width=0.48\textwidth]{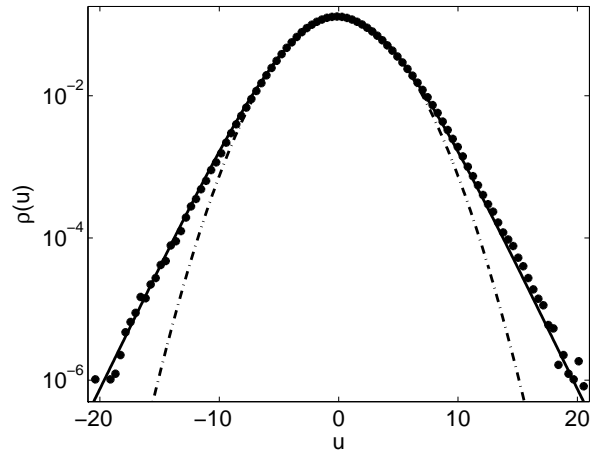}}
\caption{\label{fig:ori} Stationary distribution $\rho(u)$ of
velocity differences $u(t)$ as measured in the experiment of
Swinney et al. \cite{data_sw} at Reynolds number
$\textrm{Re}=69000$ and scale $\delta=64$. The dashed-dotted line
is a Gaussian distribution $0.1280\exp(-0.0515u^2)$, whereas the
solid line corresponds to the superstatistical formula
(\ref{rho}).}
\end{center}
\end{figure}
\begin{figure}
\begin{center}
\hfill
\parbox{0.48\textwidth}{\includegraphics[width=0.48\textwidth]{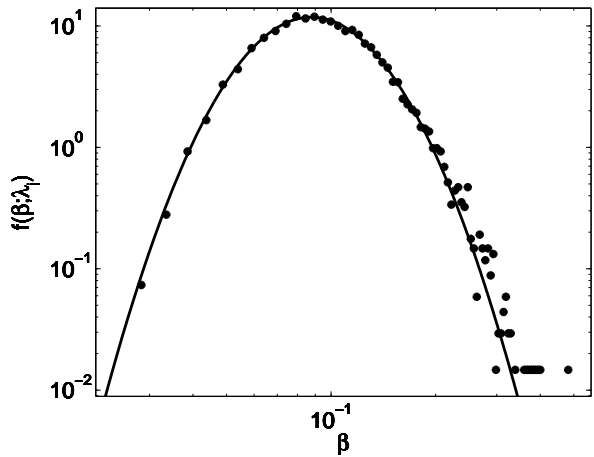}}
\caption{\label{fig:log} Example of a probability distribution
$f(\beta;\lambda_i)$ as extracted from the measured turbulent time
series of velocity differences for $\textrm{Re}=69000$ and
$\delta=64$. The solid line is a fit to the lognormal
distribution (\ref{loglam}), with $\lambda_3=3.8516$,
$\lambda_4=-2.303$ and $Z'(\lambda_i)=\sqrt{\pi/\lambda_3}$.}
\end{center}
\end{figure}
\begin{figure}
\begin{center}
\parbox{0.48\textwidth}{\includegraphics[width=0.48\textwidth]{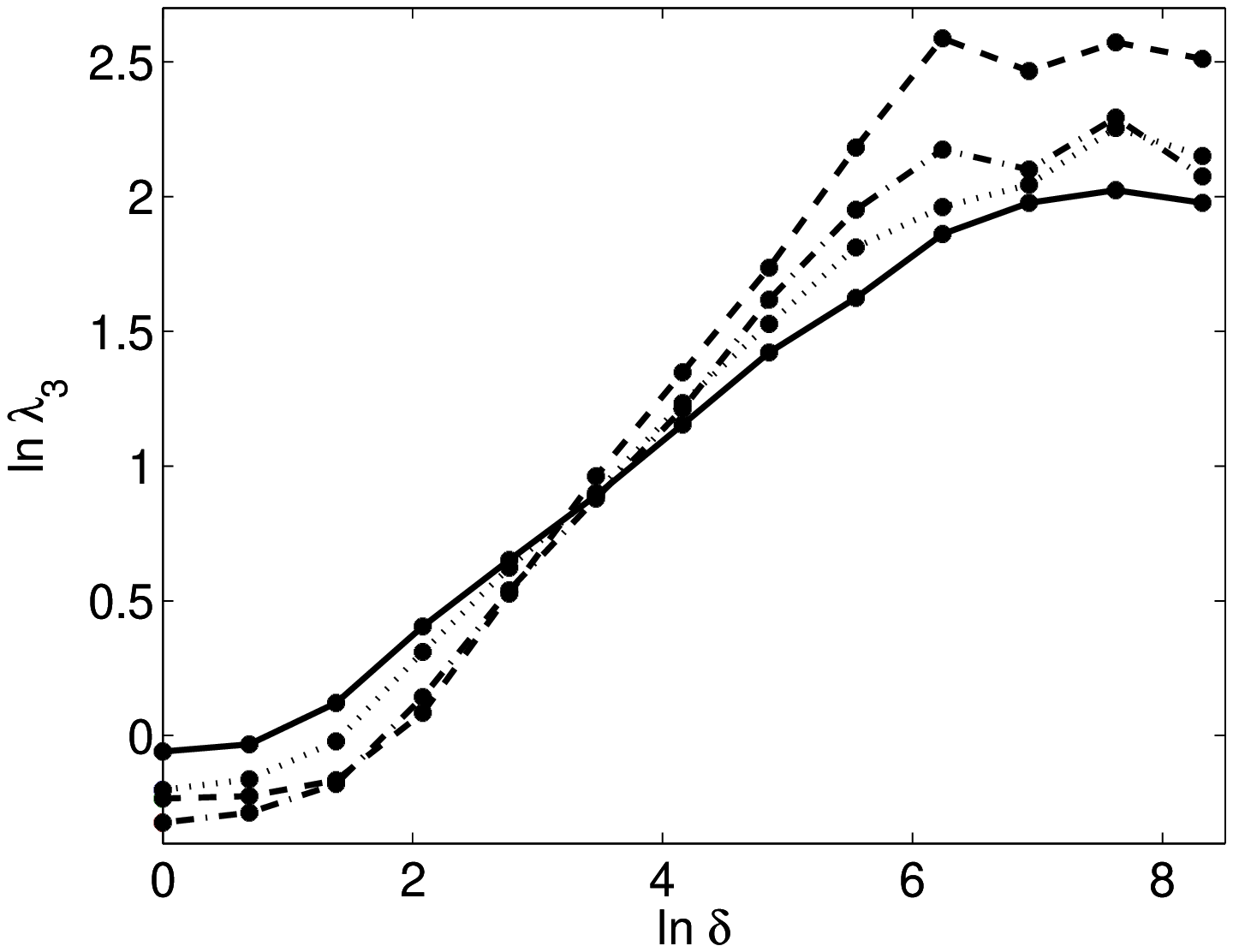}}
\hfill
\parbox{0.48\textwidth}{\includegraphics[width=0.48\textwidth]{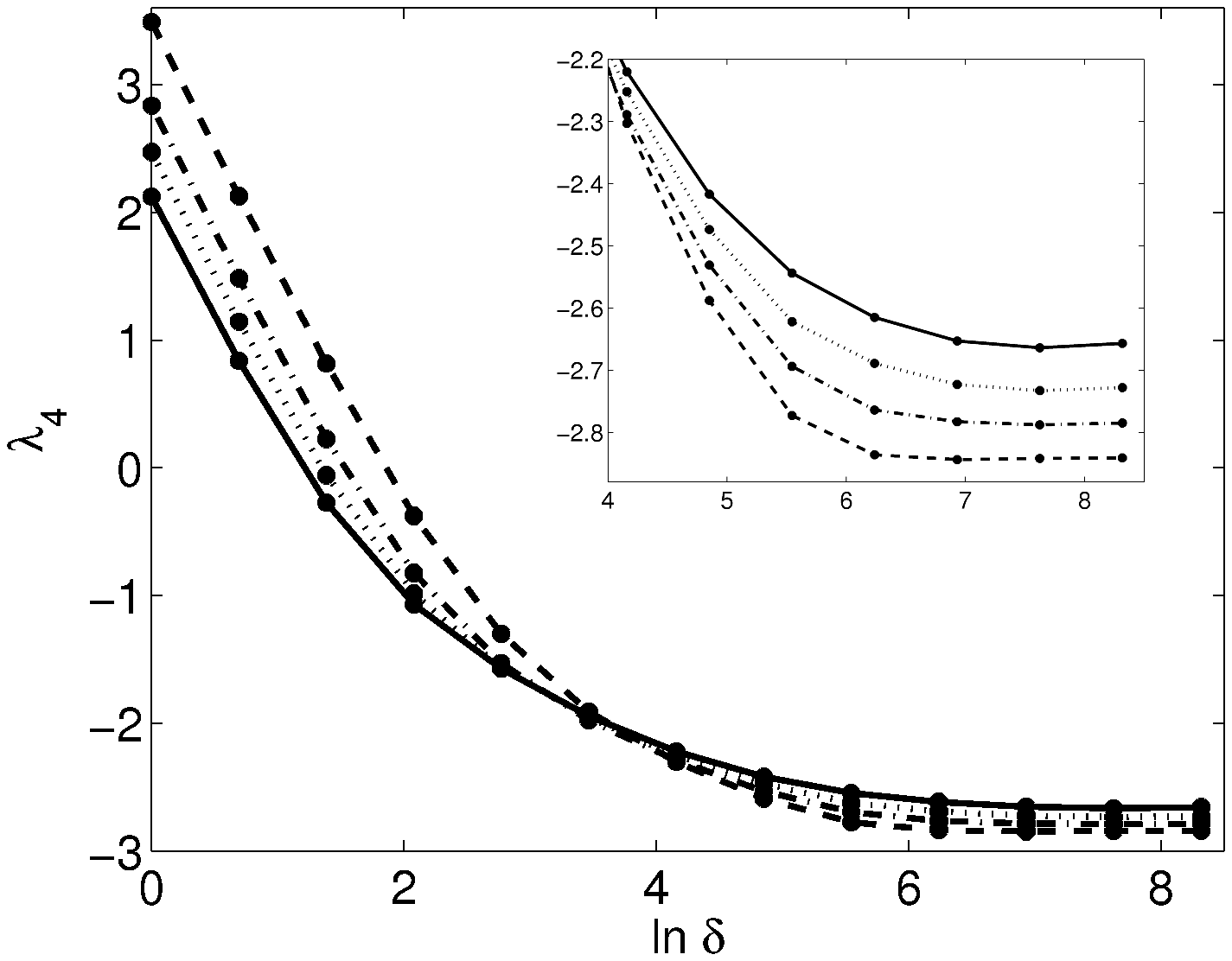}}
\caption{Dependence of the Lagrange multipliers $\lambda_3$ and
$\lambda_4$ on the scale $\delta$ and Reynolds number $\textrm{Re}$,
\label{fig:fit_para}$\textrm{Re}=540000$ (solid lines),
$\textrm{Re}=266000$ (dotted lines), $\textrm{Re}=133000$ (dashed-dotted lines) and
$\textrm{Re}=69000$ (dashed lines). The inset shows a magnification for
large values of $\delta$.}
\end{center}
\end{figure}
As a final example we now apply our methods to a complex system
that is not analytically solvable anymore: Turbulent
Taylor-Couette flow. Ideally, for a superstatistical statistical
mechanics description of this system, one would measure the set of
all positions and velocities of a large number of test particles
in the flow. This is not possible and hence, as in previous papers
\cite{unicla}, we restrict ourselves to the information that one
can get out of a scalar time series, a single measured velocity
component $v(t)$ as a function of time $t$. We use data from an
experiment performed by Lewis and Swinney \cite{data_sw}. The
stationary probability distribution $\rho (u)$ of the velocity
difference $u(t)=v(t+\delta)-v(t)$ at a given scale $\delta$
is well-known to exhibit non-Gaussian behavior, see Fig.~\ref{fig:ori} for an
example.

It has been previously shown that superstatistical techniques can
be successfully used to model the statistics of turbulent velocity
fluctuations \cite{reynolds, beck03, unicla, prl}. For a measured
time series $u(t)$, the parameter $\beta$ simply corresponds to a
local inverse variance of the measured signal, and the `cells' of
the superstatistics approach correspond to time slices of a
suitable length where this variance is measured. The turbulent
velocities are well approximated by the model of a classical ideal
superstatistical gas, meaning that for certain time intervals the
signal is Gaussian with a given variance, then it changes to
another Gaussian with a different variance, and so on. The
validity of the above approximation and the necessary time scale
separation has been checked in a previous paper \cite{unicla}. In
that paper also a general method was introduced how to to extract
the relevant time slicing (the superstatistical cell size) and
how to extract the distributions $f(\beta;\lambda_i)$ from the
signal. We do not describe this here in detail, but refer to the
paper \cite{unicla}. Using these techniques, we determined the distribution
$f(\beta;\lambda_i)$ from the measured time series, using the
experimental data of Swinney et al. for various scales $\delta$
and Reynolds numbers $\textrm{Re}$. In all cases, a lognormal distribution
turns out to be a reasonable fit for the experimentally observed
distribution $f(\beta;\lambda_i)$, see Fig.~\ref{fig:log} for an example. However, the parameters of this
lognormal distribution depend on $\delta$ and $\textrm{Re}$ in a
nontrivial way. Our results are summarized in Fig.~\ref{fig:fit_para}.

The relevance of lognormal distributions is to be expected due to
the multiplicative random processes underlying the fluctuating
energy dissipation in turbulent flows. In other words, the cascade
picture of turbulence suggests that the constraint $g(\beta)$ in
the generalized maximum entropy principle should be of the form
$g(\beta)=(\ln \beta)^2$, leading to lognormal distributions.
More surprising is the fact that our data analysis indicates that there is
a distinguished scale $\ln \delta^* \approx 3.5$, or $\delta^*
\approx 32$, where the obtained fitting parameters $\lambda_3,
\lambda_4$ are independent of Reynolds number. For $\delta <
\delta^*$, $\lambda_3$ increases with increasing Reynolds number,
whereas for $\delta > \delta^*$ it decreases. $\lambda_4$ shows
the opposite behavior, it decreases with $\textrm{Re}$ for $\delta <
\delta^*$ and increases for $\delta >\delta^*$.

One may check the quality of the superstatistical model
approximation by numerically evaluating the distribution
\cite{unicla}
\begin{equation}
\rho(u)\approx \int_0^\infty f(\beta;\lambda_i)
\sqrt{\frac{\beta}{2\pi}} e^{-\frac{1}{2}\beta u^2}d\beta
\label{rho}
\end{equation}
and comparing it with the measured stationary distribution
$\rho(u)$. Here $f(\beta;\lambda_i)$ is a lognormal distribution
with parameters as given in Fig.~\ref{fig:fit_para}. The solid line in Fig.~\ref{fig:ori}
shows this curve (\ref{rho}) for the example $\textrm{Re}=69000,
\delta=64$. Clearly, there is an excellent agreement between the
experimentally measured distribution $\rho (u)$ and the
superstatistical approximation (\ref{rho}).

Our turbulence example illustrates that the Lagrange multipliers
in the generalized entropy maximization principle, $\lambda_3$
and $\lambda_4$, do have physical meaning. Under different
conditions, in our case fixed by the scale $\delta$ under
consideration as well as the Reynolds number of the flow, these
intensive parameters have different values (see Fig.~\ref{fig:fit_para}).  In fact,
one could go as far as regarding the results in Fig.~\ref{fig:fit_para} to
describe a kind of 'equation of state' of the turbulent
Taylor-Couette flow, providing the dependence of the intensive
parameters $\lambda_3$ and $\lambda_4$ on given parameters of the
flow pattern, such as scale $\delta$ and $\textrm{Re}$. All this
illustrates that the generalized maximum entropy principles
developed in this paper are not only a mathematical exercise, but
of true physical relevance for a variety of classes of complex
systems, when only a certain limited amount of information on the
system is available.

\section{Conclusion}
In this paper we developed a maximum entropy principle for
superstatistical systems of various kinds. This principle allows
for the determination of the superstatistical distribution
$f(\beta;\lambda_i)$ of the fluctuating intensive parameter
$\beta$, given some prior information on the complex system under
consideration. Our formalism further develops previous work of
Abe et al., Crooks, and Naudts, and contains physically relevant
superstatistical universality classes, such as lognormal
superstatistics, gamma superstatistics and inverse
gamma superstatistics, as special cases. We dealt with 3
important physical examples, the superstatistical quantum harmonic
oscillator, the superstatistical classical ideal gas and time series as
generated by a turbulent Taylor-Couette flow. For the quantum
case, a new single-peaked distribution $f(\beta;\lambda_i)$ as
displayed in Fig.~\ref{fig:plot_harm_osci} arises quite naturally out of our maximum
entropy approach, whose physical relevance can be checked in
future experiments. For classical systems, other types of
distributions are relevant, such as the lognormal distribution for
turbulent flows, as displayed in Fig.~\ref{fig:log}. Our approach is a
further step to arrive at a generalized statistical formalism
relevant for large classes of complex systems with time scale
separation.


\begin{thebibliography}{99}
\bibitem{beck-cohen} C. Beck and E.G.D. Cohen, {\em Superstatistics}, Physica A {\bf 322}, 267 (2003)
\bibitem{unicla} C. Beck, E.G.D. Cohen, and H.L. Swinney, {\em From time series to superstatistics},
Phs. Rev. E {\bf 72}, 026304 (2005)
\bibitem{supergen} C. Beck and E.G.D. Cohen, {\em Superstatistical
generalization of the work fluctuation theorem}, Physica A {\bf 344},
393 (2004)
\bibitem{touchette} H. Touchette and C. Beck,
{\em Asymptotics of Superstatistics}, Phys. Rev. E {\bf 71},
016131 (2005)
\bibitem{souza} C. Tsallis and A.M.C. Souza, {\em
Constructing a statistical mechanics for Beck-Cohen superstatistics},
Phys. Rev. E {\bf 67},
026106 (2003)
\bibitem{rajagopal} A.K. Rajagopal, {\em Superstatistics -- a quantum generalization},
cond-mat/0608679
\bibitem{jizba} P. Jizba, H. Kleinert, {\em Superpositions of probability distributions},
arXiv:0802.0695
\bibitem{plastino} C. Vignat, A. Plastino and A.R. Plastino, {\em Superstatistics
based on the microcanonical ensemble},
cond-mat/0505580
\bibitem{chavanis} P.-H. Chavanis, {\em Coarse grained distributions
and superstatistics}, Physica A {\bf 359}, 177 (2006)
\bibitem{wilk} G. Wilk and Z. Wlodarczyk, {\em Interpretation of the
nonextensivity parameter $q$ in some applications of Tsallis statistics
and Levy distributions}, Phys. Rev. Lett. {\bf 84}, 2770 (2000)
\bibitem{prl01} C. Beck, {\em Dynamical foundations of nonextensive
statistical mechanics}, Phys. Rev. Lett. {\bf 87}, 180601 (2001)
\bibitem{sattin} F. Sattin, {\em Superstatistics from a different viewpoint},
Physica A {\bf 338}, 437 (2004)
\bibitem{daniels} K.~E. Daniels, C. Beck, and E. Bodenschatz,
{\em Generalized statistical mechanics and defect turbulence},
Physica D {\bf 193}, 208 (2004)
%%10
\bibitem{cosmic} C. Beck, {\em Generalized statistical mechanics
of cosmic rays}, Physica A {\bf 331}, 173 (2004)
\bibitem{maya} M. Baiesi, M. Paczuski and A.L. Stella, {\em Intensity
thresholds and the statistics of temporal occurence of solar flares},
Phys. Rev. Lett. {\bf 96}, 051103 (2006)
\bibitem{eco} Y. Ohtaki and H.H. Hasegawa, {\em Superstatistics in econophysics},
cond-mat/0312568
\bibitem{RMT} A.Y. Abul-Magd, {\em Superstatistics in random matrix theory},
Physica A {\bf 361}, 41 (2006)
\bibitem{rapisarda} S. Rizzo and A. Rapisarda, {\em
Environmental atmospheric turbulence at Florence airport},
Proceedings of the 8th Experimental Chaos Conference, Florence,
AIP Conf. Proc. {\bf 742}, 176 (2004) (cond-mat/0406684)
\bibitem{porpo} A. Porporato, G. Vico, and P.A. Fay,
{\em Superstatistics in hydro-climatic fluctuations and interannual
ecosystem productivity}, Geophys. Res. Lett. {\bf 33}, L15402 (2006)
\bibitem{reynolds} A. Reynolds, {\em
Superstatistical mechanics of tracer-particle motions
in turbulence}, Phys. Rev. Lett. {\bf 91}, 084503 (2003)
\bibitem{vKa} N.G. van Kampen, {\em Stochastic processes in physics and chemistry},
 North-Holland, London (1982)
\bibitem{heston} S.L. Heston, {\em A closed-form solution
for options with stochastic volatility with applications to bond
and currency options}, Rev. Fin. Studies {\bf 6}, 327 (1993)
\bibitem{kubo} R. Kubo, M. Toda and N. Hashitsume, {\em Statistical Physics II:
Nonequilibrium statistical mechanics}, Springer, New York (1995)
\bibitem{feller} W. Feller, {\em An introduction to probability theory and its applications, Vol. II},
John Wiley, London (1966)
\bibitem{beck03} C. Beck, {\em Lagrangian acceleration statistics
in turbulent flows}, Europhys. Lett. {\bf 64}, 151 (2003)
\bibitem{prl} C. Beck, {\em Statistics of 3-dimensional Lagrangian turbulence},
Phys. Rev. Lett. {\bf 98}, 064502 (2007)
\bibitem{ausloos} M. Ausloos and K. Ivanova, {\em
Dynamical model and nonextensive statistical mechanics
of a market index on large time windows}, Phys. Rev. E {\bf 68},
046122 (2003)
\bibitem{bouchard} J.-P. Bouchard and M. Potters,
{\em Theory of Financial Risk and Derivative Pricing}, Cambridge
University Press, Cambridge (2003)
\bibitem{eco2} H. Aoyama et al., {\em Productivity dispersion: Facts, theory, and implications},
arXiv:0805.2792
\bibitem{abul-magd2} A.Y. Abul-Magd, B. Dietz, T. Friedrich, A. Richer, {\em
Spectral fluctuations of billiards with mixed dynamics: from time series to superstatistics}, Phys. Rev. E {\bf 77}, 046202 (2008)
\bibitem{abe-thurner} S. Abe and S. Thurner, {\em
Complex networks arising from fluctuating random graphs}, Phys. Rev. E {\bf 72}, 036102 (2005)
\bibitem{queiros} S\'ilvio M. Duarte Queir\'os, {\em On Superstatistical Multiplicative-Noise Processes}, Braz. J. Phys. {\bf 38}, 203 (2008)
\bibitem{briggs} K. Briggs, C. Beck, {\em Modelling train dealys with q-exponential functions},
Physica A {\bf 378}, 498 (2007)
\bibitem{chen} L. Leon Chen, C. Beck, {\em A superstatistical model of metastasis and cancer survival},
Physica A {\bf 387}, 3162 (2008)
\bibitem{abc} S. Abe, C. Beck and G. D. Cohen, {\it Superstatistics, thermodynamics, and fluctuations,}
Phys. Rev. E {\bf 76}, 031102 (2007)
\bibitem{crooks} G. E. Crooks, {\it Beyond Boltzmann-Gibbs statistics: Maximum entropy hyperensembles
out of equilibrium,} Phys. Rev. E {\bf 75}, 041119 (2007)
\bibitem{naudts} J. Naudts, {\it Generalised thermostatistics using hyperensembles,}
AIP Conference Proceedings {\bf 84}, 965 (2007)
\bibitem{jaynes} R. D. Rosenkrantz, {\it E.T. Jaynes: papers on probability, statistics and statistical physics,}
 Kluwer (1989)
\bibitem{tsallis} C. Tsallis, {\em Possible generalization of Boltzmann-Gibbs statistics},
J. Stat. Phys. {\bf 52}, 479 (1988)
\bibitem{data_sw} G. S. Lewis and H. L. Swinney, {\it Velocity structure functions, scaling, and
transitions in high-Reynolds-number Couette-Taylor flow,} Phys. Rev. E {\bf 59}, 5457 (1999)

\end{thebibliography}
\end{document}